\documentclass[aps,prl,twocolumn,showpacs,preprintnumbers,superscriptaddress]{revtex4}
\usepackage{bm,amsmath,amssymb,latexsym,mathrsfs,enumerate,color}
\usepackage[mathcal]{euscript}
\usepackage{graphicx}
\renewcommand{\vec}[1]{\bm{#1}}
\begin{document}

\title{Controllable switching of vortex chirality in magnetic nanodisks
by a field pulse}

\author{Yuri Gaididei}
 \affiliation{Institute for Theoretical Physics, 03143 Kiev, Ukraine}

\author{Denis D. Sheka}
 \email[Corresponding author. Electronic address:\\]{denis\_sheka@univ.kiev.ua}
 \affiliation{National Taras Shevchenko University of Kiev, 03127 Kiev, Ukraine}

\author{Franz G.~Mertens}
 \affiliation{Physics Institute, University of Bayreuth, 95440 Bayreuth, Germany}

\date{October 23, 2007}

%
%

\begin{abstract}
We propose a way of fast switching the chirality in a magnetic nanodisk by
applying a field pulse. To break the symmetry with respect to clockwise or
counterclockwise chirality a mask is added by which an inhomogeneous field
influences the vortex state of a nanodisk. Using numerical spin--lattice
simulations we demonstrate that chirality can be controllably switched by a
field pulse, whose intensity is above some critical value. A mathematical
definition for the chirality of an arbitrary shaped particle is proposed.
\end{abstract}




\maketitle

In the last few years the nonlinear dynamics of magnetic nanostructures has
been extensively studied in order to develop useful devices for storage and
transmission of information. Magnetic nanodots with their vortex ground state
show a considerable promise as candidates for high density magnetic storage
and high speed magnetic random access memory \cite{Cowburn02}. The vortex
state disks are characterized by the following conserved quantities, which can
be associated with a bit of information: the \emph{polarity}, the sense of the
vortex core magnetization direction (up or down) and the \emph{chirality} or
handedness, the sense of magnetization rotation (clockwise or
counterclockwise). Until recently the main efforts went into the control of
vortex polarity and vortex motion inside the nanodots. The polarity switching
under the action of an ac magnetic field was predicted theoretically
\cite{Gaididei00} and was recently observed experimentally in Permalloy
nanodisks by using short bursts of an ac magnetic field \cite{Waeyenberge06}.
The polarity switching under the action of a dc electrical current was
predicted theoretically \cite{Caputo07} and observed experimentally quite
recently \cite{Yamada07}. The main features of the polarity switching
mechanism were clarified in Refs.~\cite{Waeyenberge06,Xiao06,Hertel07,Lee07,
Kravchuk07b,Sheka07}, where it was shown that no matter whether the switching
is induced by a magnetic field or by an electrical current it involves
creation of a vortex-antivortex pair and subsequent annihilation of the
new-born anti-vortex with the original vortex.

In standard experiments with in-plane field and circular nanodisks the vortex
chirality is not under control because of the high symmetry of the system
\cite{Schneider00}. However, the chirality is known to be controlled by
introducing an asymmetry in the particle shape, like flat edges of the disk
\cite{Schneider01,Kimura07}, rings with asymmetrical inner holes
\cite{Subramani06}, and elliptical dots \cite{Vavassori04,Mironov07}. Quite
recently it was demonstrated \cite{Choi07a} that the vortex chirality can be
switched by injecting a current pulse with appropriate amplitude, polarity and
duration. This was achieved in a multilayer structure with two different
magnetic layers, each being in a vortex state.

\begin{figure}
\includegraphics[width=0.5\columnwidth]{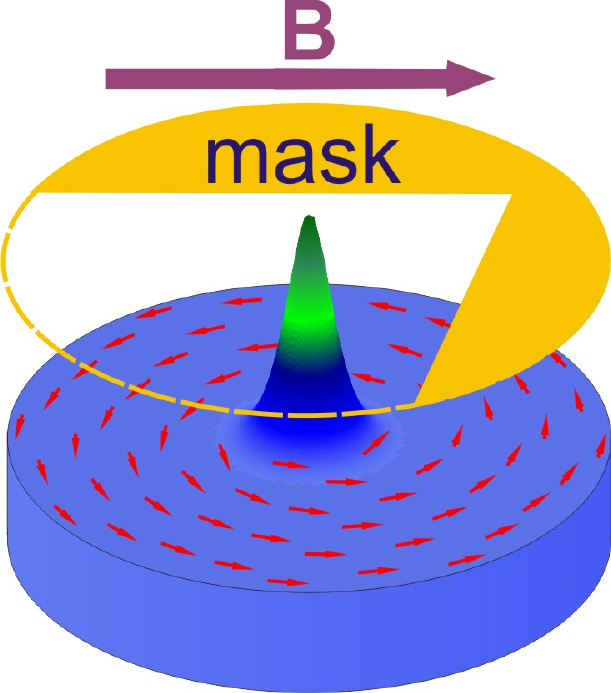}
\caption{(Color online) Schematic of the structure. The mask is added
to create an inhomogeneous magnetic field, which breaks the symmetry and
separates one chirality direction.}%
\label{fig:str}%
\end{figure}

The aim of the present Letter is to show that the vortex chirality can be
controllably switched by applying a magnetic field pulse in a circular
magnetic disk. We use the disk--shaped particle in a vortex state, see
Fig.~\ref{fig:str}. A magnetic field is applied in the disk plane, which moves
the vortex away from the disk center. If the homogeneous DC field exceeds the
value $b_{\text{an}}$, the vortex is annihilated, and finally the vortex state
switches to the monodomain state \cite{Hubert98}. After that the field is
switched off, and a vortex enters the system, however the chirality of the
new-born vortex could be either clockwise or counterclockwise
\cite{Schneider00}. To control the chirality of nucleated vortex one needs to
break the symmetry with respect to chirality. We propose to break it by
inhomogeneous field, which can be simply realized, e.g. by using a mask, which
shields a part of the disk from the field influence, see Fig.~\ref{fig:str}.

\begin{figure*}
\includegraphics[width=\textwidth]{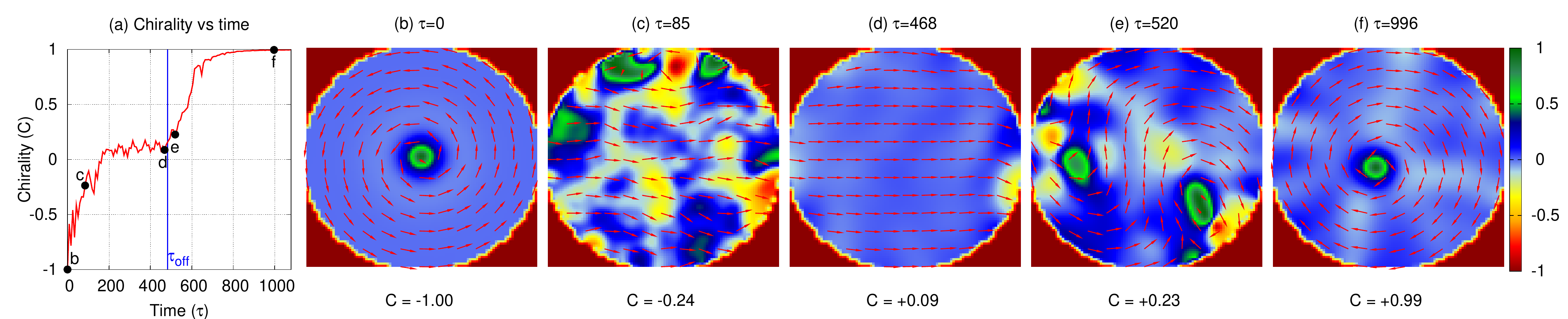}
\caption{(Color online) Time evolution of the vortex chirality switching
process from simulations with a magnetic field pulse of amplitude $b=0.2$ and
duration $\tau=480$. (a) The vortex chirality as a function of time. (b)-(f)
Snapshot images illustrating in-plane magnetization distributions at different
stages of the vortex dynamics.}
\label{fig:spin_details}%
\end{figure*}

To realize this program let us consider the dynamics of classical lattice
spins, described by the Landau--Lifshitz--Gilbert equations
\begin{equation} \label{eq:LLS-discrete}
\frac{\mathrm d \vec{S}_{\vec{n}}}{\mathrm d \tau}
= -\vec{S}_{\vec{n}}\times \frac{\partial
\mathcal{H}}{\partial \vec{S}_{\vec{n}}} - \alpha \vec{S}_{\vec{n}}
\times \frac{\mathrm d \vec{S}_{\vec{n}}}{\mathrm d \tau},
\end{equation}
where $\vec{S}_{\vec{n}}$ is a unit vector which determines the spin direction
at the lattice point $\vec{n}$, the lattice constant is chosen as the unity
length. The dimensionless time $\tau=\omega_0 t$ with $\omega_0 = 4\pi\gamma
M_S$ and $M_S$ being the saturation magnetization; $\alpha\ll1$ is a damping
coefficient. The Hamiltonian $\mathcal{H}$:
\begin{equation} \label{eq:H}
\begin{split}
\mathcal{H} &= -\frac{\ell^2}{2}\!\! \sum_{\left(\vec{n},\vec{\delta}\right)}\!
\vec{S}_{\vec{n}}\cdot \vec{S}_{\vec{n}+\vec{\delta}}
- \sum_{\vec{n}}\vec{b}\cdot \vec{S}_{\vec{n}}\\
& + \frac{1}{8\pi}\!\! \sum_{\substack{\vec{n}, \vec{n}'\\\vec{n} \neq \vec{n}'}}\!
\frac{\vec{S}_{\vec{n}}\cdot \vec{S}_{\vec{n}'}-3
\left(\vec{S}_{\vec{n}}\cdot \vec{e}_{\vec{n}\vec{n}'} \right)
\left(\vec{S}_{\vec{n}'}\cdot \vec{e}_{\vec{n}\vec{n}'} \right)}{
|\vec{n}- \vec{n}'|^3}
\end{split}
\end{equation}
consists of exchange, Zeeman and dipolar terms. Here $\ell = \sqrt{A/(4\pi
M_S^2)}$ is the exchange length ($A$ is the exchange constant), the vector
$\delta$ connects nearest neighbors, and $\vec{e}_{\vec{n}\vec{n}'} \equiv
(\vec{n} - \vec{n}')/|\vec{n} - \vec{n}'|$ is a unit vector. The dimensionless
magnetic dc field $\vec{b} = \vec{B}/4\pi M_S$.

In the continuum description, the spin dynamics is described by a
magnetization unit vector $\vec{m} = -\langle \vec{S}_{\vec{n}} \rangle$.
Usually the chirality is introduced as the sense of a magnetization rotation:
either clockwise or counterclockwise around some direction $\vec{N}$. To
describe the chirality as a continuous quantity we introduce the following
definition:
\begin{equation} \label{eq:chirality}
C = \frac{1}{L}\int_\sigma \left[\vec{\nabla}\times \vec{m}\right]
\cdot \mathrm{d}\vec{\sigma} = \frac{1}{L}\int_{\partial \sigma}\vec{m}
\cdot \mathrm{d}\vec{\tau},
\end{equation}
where $\vec{\sigma}=\sigma \vec{e}_N$ is a surface, perpendicular to
$\vec{e}_N$, and $L$ is the length of a contour, which bounds a surface
$\sigma$. For a disk-shaped particle of radius $R$ with in-plane magnetization
distribution, described by the angular variable $\phi=\arctan(m^y/m^x)$, the
chirality $C = \frac{1}{2\pi}\int_0^{2\pi}\sin(\phi-\chi)\mathrm{d}\chi$
around the circle of radius $R$, where $\chi$ is a polar angle in the disk
plane. The in-plane magnetization distribution in the vortex state nanodisk is
described by the formula $\phi = \chi +C\pi/2$.

The physical idea of the chirality switching is very simple. In an external dc
magnetic field $b$, which is stronger that some annihilation field
$b_{\text{an}}$, but below the saturation field $b_{\text{sat}}$, the
monodomain state appears. However, if the field $b$ is inhomogeneous, the
monodomain state is nonuniform. By choosing the shape of the field, one can
create this intermediate monodomain state with a certain nonzero chirality.
After that, if we switch off the field, the vortex will be nucleated with a
predefined chirality.

To study the vortex dynamics, we have performed numerical simulations of the
discrete spin-lattice Eqs.~\eqref{eq:LLS-discrete}. We consider the case of
thin nanodots where the magnetization does not depend on the z-coordinate, the
details of the method are described in Ref.~\cite{Caputo07a}. We have
integrated numerically the set of Eqs.~\eqref{eq:LLS-discrete} over a square
lattice using the fourth-order Runge-Kutta scheme with time step $\Delta\tau =
0.01$. The lattice is bounded by a circle of diameter $2R$. In most of the
simulations the system diameter $2R=50$, the thickness $h=10$, other
parameters: $\ell = 1.33$, $\alpha =0.01$. We applied an inhomogeneous field
along the $x$ axis by the shape $b(x,y) = b\Theta(\xi-x)\Theta(\xi-y)$ with
$\Theta(x)$ being the Heaviside step function. We choose $\xi/(2R)=0.85$. Such
a field can be realized by a mask which covers a part of the disk, see
Fig.~\ref{fig:str}.

\begin{figure}
\includegraphics[width=\columnwidth]{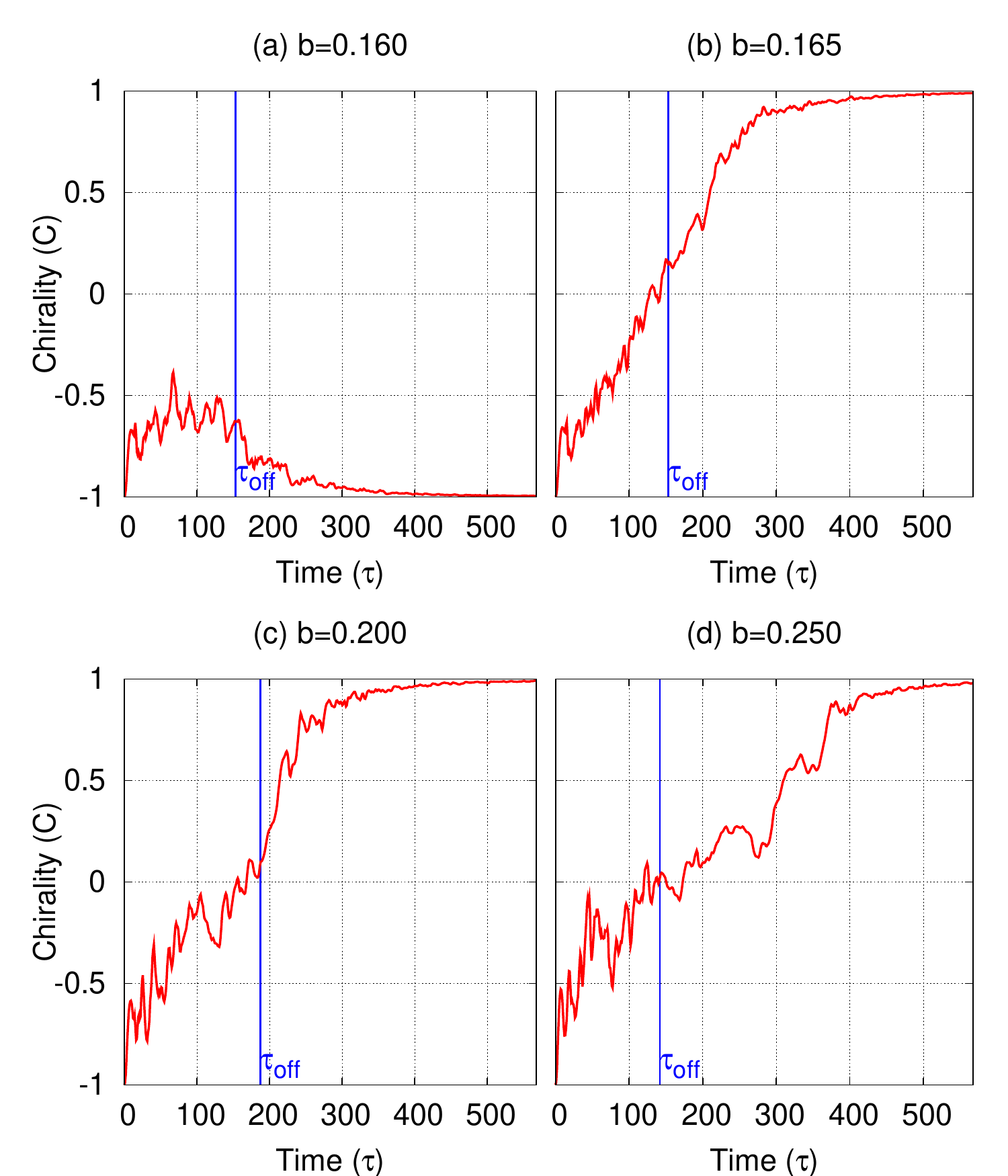}
\caption{(Color online) Time dependence of the vortex chirality for
different amplitudes of field pulses: (a) without switching,
(b)-(d) with chirality switching. The switching occurs, when the field amplitude
exceeds the critical value $b_{\text{an}}\approx0.163.$
}
\label{fig:chi_vs_time}%
\end{figure}

The temporal evolution of the chirality is presented in
Fig.~\ref{fig:spin_details}(a). Initially the chirality was $C=-1$
(counterclockwise), see Fig.~\ref{fig:spin_details}(b). Under the action of
the field pulse the vortex is pushed away in the direction of the north pole
of the disk, which is accompanied by a decrease of the absolute value of the
chirality, see Fig.~\ref{fig:spin_details}(c). The next monodomain stage is
presented in Fig.~\ref{fig:spin_details}(d). The vortex is already
annihilated, but the total chirality takes a nonzero (positive) value due to
the mask influence. After that the field pulse is switched off, and the new
vortex enters the disk from the south pole with a predefined (positive)
chirality, see Fig.~\ref{fig:spin_details}(e). This vortex quickly moves to
the disk center, and finally the chirality is $C=+1$ (clockwise), see
Fig.~\ref{fig:spin_details}(f). The switching process takes place in a wide
range of parameters of the field pulse, when the field intensity
$b>b_{\text{an}}$, see Fig.~\ref{fig:chi_vs_time}.

Using typical parameters for permalloy disks \footnote{For the Py particle the
exchange constant $A=2.6 \mu$erg/cm, the saturation magnetization $M_S=0.86$
kG.}, one can estimate the exchange length $\ell = 5.3$ nm, which corresponds
to a particle with $2R=200$ nm and $h=40$ nm. The typical time unit
$1/\omega_0=50$ ps and the field unit $4\pi M_S =10$ kOe. The typical
switching field intensity is about 140 Oe and the typical switching time is
about 7 ns.

To summarize, we have studied the chirality switching under the action of a dc
magnetic field. In contrast to previous studies on the chirality control, we
showed that the switching can be managed in a pure disk nanoparticle: to break
the symmetry the usage of the mask is proposed. Usually the chirality is
considered as either clockwise or counterclockwise. By considering the
chirality as a continuous quantity, we were able to control the switching
process: the key point is an intermediate state with a nonzero chirality. We
consider here only one possible realization of the chirality switching with a
very simple mask. We expect that the usage of a mask will be prominent for
different applications of nonlinear magnetization dynamics in nanoparticles.

\acknowledgments

The authors thank H.~Stoll and M.~F\"ahnle for stimulating discussions. D.S.,
Yu.G. thank the University of Bayreuth, where this work was performed, for
kind hospitality, and acknowledge the support from DLR grant No.~UKR~05/055.
D.S. acknowledges the support from the Alexander von Humboldt--Foundation and
the grant No.~F25.2/098 from the Fundamental Researches State Fund of Ukraine.

\begin{thebibliography}{19}
\expandafter\ifx\csname natexlab\endcsname\relax\def\natexlab#1{#1}\fi
\expandafter\ifx\csname bibnamefont\endcsname\relax
  \def\bibnamefont#1{#1}\fi
\expandafter\ifx\csname bibfnamefont\endcsname\relax
  \def\bibfnamefont#1{#1}\fi
\expandafter\ifx\csname citenamefont\endcsname\relax
  \def\citenamefont#1{#1}\fi
\expandafter\ifx\csname url\endcsname\relax
  \def\url#1{\texttt{#1}}\fi
\expandafter\ifx\csname urlprefix\endcsname\relax\def\urlprefix{URL }\fi
\providecommand{\bibinfo}[2]{#2}
\providecommand{\eprint}[2][]{\url{#2}}

\bibitem[{\citenamefont{Cowburn}(2002)}]{Cowburn02}
\bibinfo{author}{\bibfnamefont{R.~P.} \bibnamefont{Cowburn}},
  \bibinfo{journal}{J.~Magn. Magn. Mater.} \textbf{\bibinfo{volume}{242-245}},
  \bibinfo{pages}{505} (\bibinfo{year}{2002}),
  \urlprefix\url{http://dx.doi.org/10.1016/S0304-8853(01)01086-1}.

\bibitem[{\citenamefont{Gaididei et~al.}(2000)\citenamefont{Gaididei,
  Kamppeter, Mertens, and Bishop}}]{Gaididei00}
\bibinfo{author}{\bibfnamefont{Y.}~\bibnamefont{Gaididei}},
  \bibinfo{author}{\bibfnamefont{T.}~\bibnamefont{Kamppeter}},
  \bibinfo{author}{\bibfnamefont{F.~G.} \bibnamefont{Mertens}},
  \bibnamefont{and} \bibinfo{author}{\bibfnamefont{A.~R.}
  \bibnamefont{Bishop}}, \bibinfo{journal}{Phys. Rev. B}
  \textbf{\bibinfo{volume}{61}}, \bibinfo{pages}{9449} (\bibinfo{year}{2000}),
  \urlprefix\url{http://link.aps.org/abstract/PRB/v61/p9449}.

\bibitem[{\citenamefont{Waeyenberge et~al.}(2006)\citenamefont{Waeyenberge,
  Puzic, Stoll, Chou, Tyliszczak, Hertel, Fahnle, Bruckl, Rott, Reiss
  et~al.}}]{Waeyenberge06}
\bibinfo{author}{\bibfnamefont{V.~B.} \bibnamefont{Waeyenberge}},
  \bibinfo{author}{\bibfnamefont{A.}~\bibnamefont{Puzic}},
  \bibinfo{author}{\bibfnamefont{H.}~\bibnamefont{Stoll}},
  \bibinfo{author}{\bibfnamefont{K.~W.} \bibnamefont{Chou}},
  \bibinfo{author}{\bibfnamefont{T.}~\bibnamefont{Tyliszczak}},
  \bibinfo{author}{\bibfnamefont{R.}~\bibnamefont{Hertel}},
  \bibinfo{author}{\bibfnamefont{M.}~\bibnamefont{Fahnle}},
  \bibinfo{author}{\bibfnamefont{H.}~\bibnamefont{Bruckl}},
  \bibinfo{author}{\bibfnamefont{K.}~\bibnamefont{Rott}},
  \bibinfo{author}{\bibfnamefont{G.}~\bibnamefont{Reiss}},
  \bibnamefont{et~al.}, \bibinfo{journal}{Nature}
  \textbf{\bibinfo{volume}{444}}, \bibinfo{pages}{461} (\bibinfo{year}{2006}),
  ISSN \bibinfo{issn}{0028-0836},
  \urlprefix\url{http://dx.doi.org/10.1038/nature05240}.

\bibitem[{\citenamefont{Caputo et~al.}(2007{\natexlab{a}})\citenamefont{Caputo,
  Gaididei, Mertens, and Sheka}}]{Caputo07}
\bibinfo{author}{\bibfnamefont{J.-G.} \bibnamefont{Caputo}},
  \bibinfo{author}{\bibfnamefont{Y.}~\bibnamefont{Gaididei}},
  \bibinfo{author}{\bibfnamefont{F.~G.} \bibnamefont{Mertens}},
  \bibnamefont{and} \bibinfo{author}{\bibfnamefont{D.~D.} \bibnamefont{Sheka}},
  \bibinfo{journal}{Phys. Rev. Lett.} \textbf{\bibinfo{volume}{98}},
  \bibinfo{eid}{056604} (pages~\bibinfo{numpages}{4})
  (\bibinfo{year}{2007}{\natexlab{a}}),
  \urlprefix\url{http://link.aps.org/abstract/PRL/v98/e056604}.

\bibitem[{\citenamefont{Yamada et~al.}(2007)\citenamefont{Yamada, Kasai,
  Nakatani, Kobayashi, Kohno, Thiaville, and Ono}}]{Yamada07}
\bibinfo{author}{\bibfnamefont{K.}~\bibnamefont{Yamada}},
  \bibinfo{author}{\bibfnamefont{S.}~\bibnamefont{Kasai}},
  \bibinfo{author}{\bibfnamefont{Y.}~\bibnamefont{Nakatani}},
  \bibinfo{author}{\bibfnamefont{K.}~\bibnamefont{Kobayashi}},
  \bibinfo{author}{\bibfnamefont{H.}~\bibnamefont{Kohno}},
  \bibinfo{author}{\bibfnamefont{A.}~\bibnamefont{Thiaville}},
  \bibnamefont{and} \bibinfo{author}{\bibfnamefont{T.}~\bibnamefont{Ono}},
  \bibinfo{journal}{Nat Mater} \textbf{\bibinfo{volume}{6}},
  \bibinfo{pages}{270} (\bibinfo{year}{2007}), ISSN \bibinfo{issn}{1476-1122},
  \urlprefix\url{http://dx.doi.org/10.1038/nmat1867}.

\bibitem[{\citenamefont{Xiao et~al.}(2006)\citenamefont{Xiao, Rudge, Choi,
  Hong, and Donohoe}}]{Xiao06}
\bibinfo{author}{\bibfnamefont{Q.~F.} \bibnamefont{Xiao}},
  \bibinfo{author}{\bibfnamefont{J.}~\bibnamefont{Rudge}},
  \bibinfo{author}{\bibfnamefont{B.~C.} \bibnamefont{Choi}},
  \bibinfo{author}{\bibfnamefont{Y.~K.} \bibnamefont{Hong}}, \bibnamefont{and}
  \bibinfo{author}{\bibfnamefont{G.}~\bibnamefont{Donohoe}},
  \bibinfo{journal}{Applied Physics Letters} \textbf{\bibinfo{volume}{89}},
  \bibinfo{eid}{262507} (pages~\bibinfo{numpages}{3}) (\bibinfo{year}{2006}),
  \urlprefix\url{http://link.aip.org/link/?APL/89/262507/1}.

\bibitem[{\citenamefont{Hertel et~al.}(2007)\citenamefont{Hertel, Gliga,
  Fahnle, and Schneider}}]{Hertel07}
\bibinfo{author}{\bibfnamefont{R.}~\bibnamefont{Hertel}},
  \bibinfo{author}{\bibfnamefont{S.}~\bibnamefont{Gliga}},
  \bibinfo{author}{\bibfnamefont{M.}~\bibnamefont{Fahnle}}, \bibnamefont{and}
  \bibinfo{author}{\bibfnamefont{C.~M.} \bibnamefont{Schneider}},
  \bibinfo{journal}{Phys. Rev. Lett.} \textbf{\bibinfo{volume}{98}},
  \bibinfo{eid}{117201} (pages~\bibinfo{numpages}{4}) (\bibinfo{year}{2007}),
  \urlprefix\url{http://link.aps.org/abstract/PRL/v98/e117201}.

\bibitem[{\citenamefont{Lee et~al.}(2007)\citenamefont{Lee, Guslienko, Lee, and
  Kim}}]{Lee07}
\bibinfo{author}{\bibfnamefont{K.-S.} \bibnamefont{Lee}},
  \bibinfo{author}{\bibfnamefont{K.~Y.} \bibnamefont{Guslienko}},
  \bibinfo{author}{\bibfnamefont{J.-Y.} \bibnamefont{Lee}}, \bibnamefont{and}
  \bibinfo{author}{\bibfnamefont{S.-K.} \bibnamefont{Kim}},
  \emph{\bibinfo{title}{Ultrafast vortex-core reversal dynamics in
  ferromagnetic nanodots}} (\bibinfo{year}{2007}), \eprint{cond-mat/0703538},
  \urlprefix\url{http://www.arxiv.org/abs/cond-mat/0703538}.

\bibitem[{\citenamefont{Kravchuk et~al.}(2007)\citenamefont{Kravchuk, Sheka,
  Gaididei, and Mertens}}]{Kravchuk07b}
\bibinfo{author}{\bibfnamefont{V.~P.} \bibnamefont{Kravchuk}},
  \bibinfo{author}{\bibfnamefont{D.~D.} \bibnamefont{Sheka}},
  \bibinfo{author}{\bibfnamefont{Y.}~\bibnamefont{Gaididei}}, \bibnamefont{and}
  \bibinfo{author}{\bibfnamefont{F.~G.} \bibnamefont{Mertens}},
  \emph{\bibinfo{title}{Controlled vortex core switching in a magnetic nanodisk
  by a rotating field}} (\bibinfo{year}{2007}), \eprint{arXiv:0705.2046},
  \urlprefix\url{http://arxiv.org/abs/0705.2046}.

\bibitem[{\citenamefont{Sheka}(2007)}]{Sheka07}
\bibinfo{author}{\bibfnamefont{D.~D.} \bibnamefont{Sheka}},
  \bibinfo{journal}{Phys. Rev. B} \textbf{\bibinfo{volume}{75}},
  \bibinfo{eid}{107401} (pages~\bibinfo{numpages}{2}) (\bibinfo{year}{2007}),
  \urlprefix\url{http://link.aps.org/abstract/PRB/v75/e107401}.

\bibitem[{\citenamefont{Schneider et~al.}(2000)\citenamefont{Schneider,
  Hoffmann, and Zweck}}]{Schneider00}
\bibinfo{author}{\bibfnamefont{M.}~\bibnamefont{Schneider}},
  \bibinfo{author}{\bibfnamefont{H.}~\bibnamefont{Hoffmann}}, \bibnamefont{and}
  \bibinfo{author}{\bibfnamefont{J.}~\bibnamefont{Zweck}},
  \bibinfo{journal}{Appl. Phys. Lett.} \textbf{\bibinfo{volume}{77}},
  \bibinfo{pages}{2909} (\bibinfo{year}{2000}),
  \urlprefix\url{http://link.aip.org/link/?APL/77/2909/1}.

\bibitem[{\citenamefont{Schneider et~al.}(2001)\citenamefont{Schneider,
  Hoffmann, and Zweck}}]{Schneider01}
\bibinfo{author}{\bibfnamefont{M.}~\bibnamefont{Schneider}},
  \bibinfo{author}{\bibfnamefont{H.}~\bibnamefont{Hoffmann}}, \bibnamefont{and}
  \bibinfo{author}{\bibfnamefont{J.}~\bibnamefont{Zweck}},
  \bibinfo{journal}{Appl. Phys. Lett.} \textbf{\bibinfo{volume}{79}},
  \bibinfo{pages}{3113} (\bibinfo{year}{2001}),
  \urlprefix\url{http://link.aip.org/link/?APL/79/3113/1}.

\bibitem[{\citenamefont{Kimura et~al.}(2007)\citenamefont{Kimura, Otani,
  Masaki, Ishida, Antos, and Shibata}}]{Kimura07}
\bibinfo{author}{\bibfnamefont{T.}~\bibnamefont{Kimura}},
  \bibinfo{author}{\bibfnamefont{Y.}~\bibnamefont{Otani}},
  \bibinfo{author}{\bibfnamefont{H.}~\bibnamefont{Masaki}},
  \bibinfo{author}{\bibfnamefont{T.}~\bibnamefont{Ishida}},
  \bibinfo{author}{\bibfnamefont{R.}~\bibnamefont{Antos}}, \bibnamefont{and}
  \bibinfo{author}{\bibfnamefont{J.}~\bibnamefont{Shibata}},
  \bibinfo{journal}{Appl. Phys. Lett.} \textbf{\bibinfo{volume}{90}},
  \bibinfo{eid}{132501} (pages~\bibinfo{numpages}{3}) (\bibinfo{year}{2007}),
  \urlprefix\url{http://link.aip.org/link/?APL/90/132501/1}.

\bibitem[{\citenamefont{Subramani et~al.}(2006)\citenamefont{Subramani,
  Geerpuram, Baskaran, Friedlund, and Metlushko}}]{Subramani06}
\bibinfo{author}{\bibfnamefont{A.}~\bibnamefont{Subramani}},
  \bibinfo{author}{\bibfnamefont{D.}~\bibnamefont{Geerpuram}},
  \bibinfo{author}{\bibfnamefont{V.}~\bibnamefont{Baskaran}},
  \bibinfo{author}{\bibfnamefont{J.}~\bibnamefont{Friedlund}},
  \bibnamefont{and}
  \bibinfo{author}{\bibfnamefont{V.}~\bibnamefont{Metlushko}},
  \bibinfo{journal}{Physica C: Superconductivity}
  \textbf{\bibinfo{volume}{437-438}}, \bibinfo{pages}{293}
  (\bibinfo{year}{2006}),
  \urlprefix\url{http://www.sciencedirect.com/science/article/B6TVJ-4J624M1-5/%
2/6a578dde6b9baa5c53e8a7655800cd16}.

\bibitem[{\citenamefont{Vavassori et~al.}(2004)\citenamefont{Vavassori,
  Zaluzec, Metlushko, Novosad, Ilic, and Grimsditch}}]{Vavassori04}
\bibinfo{author}{\bibfnamefont{P.}~\bibnamefont{Vavassori}},
  \bibinfo{author}{\bibfnamefont{N.}~\bibnamefont{Zaluzec}},
  \bibinfo{author}{\bibfnamefont{V.}~\bibnamefont{Metlushko}},
  \bibinfo{author}{\bibfnamefont{V.}~\bibnamefont{Novosad}},
  \bibinfo{author}{\bibfnamefont{B.}~\bibnamefont{Ilic}}, \bibnamefont{and}
  \bibinfo{author}{\bibfnamefont{M.}~\bibnamefont{Grimsditch}},
  \bibinfo{journal}{Phys. Rev. B} \textbf{\bibinfo{volume}{69}},
  \bibinfo{eid}{214404} (pages~\bibinfo{numpages}{6}) (\bibinfo{year}{2004}),
  \urlprefix\url{http://link.aps.org/abstract/PRB/v69/e214404}.

\bibitem[{\citenamefont{Mironov et~al.}(2007)\citenamefont{Mironov, Gribkov,
  Fraerman, Gusev, Vdovichev, Karetnikova, Nefedov, and
  Shereshevsky}}]{Mironov07}
\bibinfo{author}{\bibfnamefont{V.}~\bibnamefont{Mironov}},
  \bibinfo{author}{\bibfnamefont{B.}~\bibnamefont{Gribkov}},
  \bibinfo{author}{\bibfnamefont{A.}~\bibnamefont{Fraerman}},
  \bibinfo{author}{\bibfnamefont{S.}~\bibnamefont{Gusev}},
  \bibinfo{author}{\bibfnamefont{S.}~\bibnamefont{Vdovichev}},
  \bibinfo{author}{\bibfnamefont{I.}~\bibnamefont{Karetnikova}},
  \bibinfo{author}{\bibfnamefont{I.}~\bibnamefont{Nefedov}}, \bibnamefont{and}
  \bibinfo{author}{\bibfnamefont{I.}~\bibnamefont{Shereshevsky}},
  \bibinfo{journal}{J.~Magn. Magn. Mater.} \textbf{\bibinfo{volume}{312}},
  \bibinfo{pages}{153} (\bibinfo{year}{2007}),
  \urlprefix\url{http://www.sciencedirect.com/science/article/B6TJJ-4M6S933-1/%
2/339dcee0d40dbfe9b27888e9b44a70f9}.

\bibitem[{\citenamefont{Choi et~al.}(2007)\citenamefont{Choi, Rudge, Girgis,
  Kolthammer, Hong, and Lyle}}]{Choi07a}
\bibinfo{author}{\bibfnamefont{B.~C.} \bibnamefont{Choi}},
  \bibinfo{author}{\bibfnamefont{J.}~\bibnamefont{Rudge}},
  \bibinfo{author}{\bibfnamefont{E.}~\bibnamefont{Girgis}},
  \bibinfo{author}{\bibfnamefont{J.}~\bibnamefont{Kolthammer}},
  \bibinfo{author}{\bibfnamefont{Y.~K.} \bibnamefont{Hong}}, \bibnamefont{and}
  \bibinfo{author}{\bibfnamefont{A.}~\bibnamefont{Lyle}},
  \bibinfo{journal}{Appl. Phys. Lett.} \textbf{\bibinfo{volume}{91}},
  \bibinfo{eid}{022501} (pages~\bibinfo{numpages}{3}) (\bibinfo{year}{2007}),
  \urlprefix\url{http://link.aip.org/link/?APL/91/022501/1}.

\bibitem[{\citenamefont{Hubert and Sch{\" a}fer}(1998)}]{Hubert98}
\bibinfo{author}{\bibfnamefont{A.}~\bibnamefont{Hubert}} \bibnamefont{and}
  \bibinfo{author}{\bibfnamefont{R.}~\bibnamefont{Sch{\" a}fer}},
  \emph{\bibinfo{title}{Magnetic domains}}
  (\bibinfo{publisher}{Springer--Verlag}, \bibinfo{address}{Berlin},
  \bibinfo{year}{1998}).

\bibitem[{\citenamefont{Caputo et~al.}(2007{\natexlab{b}})\citenamefont{Caputo,
  Gaididei, Kravchuk, Mertens, and Sheka}}]{Caputo07a}
\bibinfo{author}{\bibfnamefont{J.-G.} \bibnamefont{Caputo}},
  \bibinfo{author}{\bibfnamefont{Y.}~\bibnamefont{Gaididei}},
  \bibinfo{author}{\bibfnamefont{V.~P.} \bibnamefont{Kravchuk}},
  \bibinfo{author}{\bibfnamefont{F.~G.} \bibnamefont{Mertens}},
  \bibnamefont{and} \bibinfo{author}{\bibfnamefont{D.~D.} \bibnamefont{Sheka}},
  \emph{\bibinfo{title}{Effective anisotropy of thin nanomagnets: beyond the
  surface anisotropy approach}} (\bibinfo{year}{2007}{\natexlab{b}}),
  \eprint{arXiv:0705.1555}, \urlprefix\url{http://arxiv.org/abs/0705.1555}.

\end{thebibliography}

\end{document}